\documentclass[aps, pra, superscriptaddress, floatfix, reprint, showpacs, twocolumn]{revtex4-1}

\usepackage[normalem]{ulem}
\usepackage[]{color}

\usepackage{times, amsmath, amssymb, bm, wasysym, graphicx, siunitx, hyperref} \pdfoutput=1

\newcommand{\ket}[1]{|#1\rangle} 
\newcommand{\bra}[1]{\langle#1|} 
\newcommand{\braket}[2]{\langle #1 | #2 \rangle} 
\newcommand{\ketbra}[2]{| #1 \rangle \langle #2 |}

\newcommand{\A}{A}
\newcommand{\D}{D}
\newcommand{\B}{\text{B}}

\DeclareMathOperator{\tr}{Tr}
\DeclareMathOperator{\E}{E}
\DeclareMathOperator{\Var}{Var}
\DeclareMathOperator{\Cov}{Cov}

\newcommand{\arxiv}[1]{\href{http://arxiv.org/abs/#1}{\tt arXiv:\nolinkurl{#1}}}

\begin{document}

\title{Spatial propagation of excitonic coherence enables ratcheted energy transfer}
\date{\today}
\author{Stephan Hoyer}
\affiliation{Department of Physics, University of California, Berkeley, CA 94720, USA}
\affiliation{Berkeley Quantum Information and Computation Center, University of California, Berkeley, CA 94720, USA}
\author{Akihito Ishizaki}
\altaffiliation{Present address: Department of Theoretical and Computational Molecular Science,
	Institute for Molecular Science,
	National Institutes of Natural Sciences,
	Okazaki 444-8585, Japan}
\affiliation{Department of Chemistry, University of California, Berkeley, CA 94720, USA}
\affiliation{Physical Bioscience Division, Lawrence Berkeley National Laboratory, Berkeley, CA 94720, USA}
\author{K. Birgitta Whaley}
\affiliation{Berkeley Quantum Information and Computation Center, University of California, Berkeley, CA 94720, USA}
\affiliation{Department of Chemistry, University of California, Berkeley, CA 94720, USA}

\pacs{87.15.hj, 05.60.Gg, 71.35.-y}

%TC:break Abstract
\begin{abstract}
Experimental evidence shows that a variety of photosynthetic systems can preserve quantum beats in the process of electronic energy transfer, even at room temperature.
However, whether this quantum coherence arises \emph{in vivo} and whether it has any biological function have remained unclear.
Here we present a theoretical model that suggests that the creation and recreation of coherence under natural conditions is ubiquitous.
Our model allows us to theoretically demonstrate a mechanism for a ratchet effect enabled by quantum coherence, in a design inspired by an energy transfer pathway in the Fenna-Matthews-Olson complex of the green sulfur bacteria.
This suggests a possible biological role for coherent oscillations in spatially directing energy transfer.
Our results emphasize the importance of analyzing long-range energy transfer in terms of transfer between inter-complex coupling (ICC) states rather than between site or exciton states.
\end{abstract}

\maketitle

\section{Introduction}

Mounting experimental evidence for electronic quantum coherence in photosynthetic energy transfer \cite{Savikhin1997, Engel2007, Collini2010, Panitchayangkoon2010, Calhoun2009} has spawned much debate about both the detailed nature and the biological role of such quantum dynamical features.
Quantum coherence is usually encountered in the first, light harvesting stage of photosynthesis. It includes two distinct but not mutually excludeusive phenomena that can be differentiated by the choice of basis used to describe the electronic excitations.
In the site basis, corresponding to the excitation of individual pigment molecules, coherence emerges in molecular aggregates even in thermal equilibrium, since eigenstates are delocalized over multiple chromophores.
Such coherence between sites can enhance the rate of biological energy transfer by up to an order of magnitude \cite{Scholes2000, Sumi1999, Jang2004, Hossein-Nejad2011}.
In contrast, it is coherence in the exciton basis, that is, superpositions of the energy eigenstates, which drives quantum beating via the Schr\"{o}dinger equation.
It is this type of coherence on which we will focus here, and thus in the remainder of this paper, the term `coherence' refers to coherence in the exciton basis.
Photosynthetic systems at ambient temperatures have been shown to exhibit this kind of quantum beating when artificially excited \cite{Panitchayangkoon2010, Collini2010}, but the significance of these discoveries remains unclear.
A broad deficiency is the lack of plausible physical mechanisms for how this coherence could arise in and influence biological energy transfer.
For example, the suggestion that transport in these systems features speedups reminiscent of quantum search algorithms \cite{Engel2007} has been shown to be invalid \cite{Hoyer2010, Mohseni2008}.
Experimental observations of long-lasting and delocalized quantum beats alone are not sufficient to determine that they are biologically relevant, since these features arise due to strong inter-chromophoric couplings \cite{SchlauCohen:2012wh}, which independently yield fast transport rates under almost any theoretical model, even those which entirely neglect quantum coherence.

\begin{figure}[] 
	\includegraphics[]{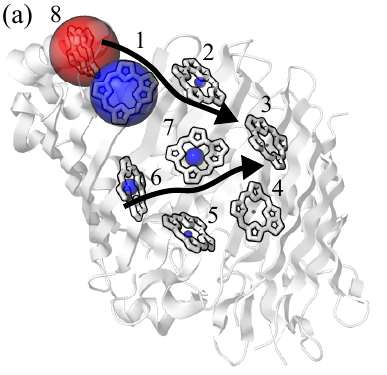}
	\includegraphics[]{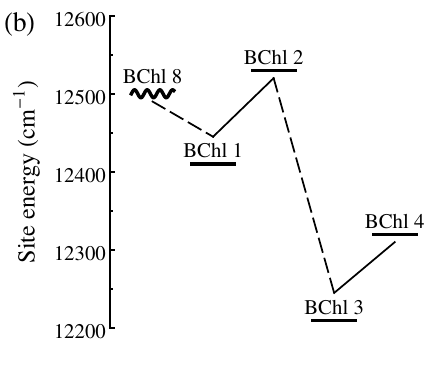}
	\caption{\label{fig:fmo-pathways}(Color online) Energy transfer pathways in a monomer of the FMO complex of \emph{Chlorobaculum tepidum} ({\it C. tepidum}). (a) Side view of a monomer of the FMO complex \cite{Fenna1975}, showing the primary energy transfer pathways \cite{Brixner2005} toward the reaction center via site 3 and the 
	inter-complex coupling (ICC) basis states that couple site 8 to the remainder of the complex. Site occupation probabilities for the ICC basis states are proportional to the area of the colored circles. (b) Site energies along the upper energy transfer pathway depicted in panel (a), with the energy of site 8 approximated by the antenna baseplate energy \cite{Ishizaki2009a}. Lines between sites indicate weak (dashed, $\SI{10}{cm^{-1}} < J < \SI{40}{cm^{-1}}$) and strong (solid, $J > \SI{40}{cm^{-1}}$) electronic couplings. Room temperature is approximately \SI{200}{cm^{-1}}.}
\end{figure}

In this work we address the question of what physical mechanisms lie behind the origin and biological role of electronic coherence in the exciton basis.  While our theoretical analysis is general, to make illustrative demonstrations of specific mechanistic features
we take physical parameters from a prototypical system, the Fenna-Matthews-Olson (FMO) complex of green sulfur bacteria. FMO is an extensively studied protein-pigment complex \cite{Adolphs2006, Blankenship2002, Fenna1975} that exhibits excitonic quantum beats \cite{Panitchayangkoon2010} and entanglement \cite{Sarovar2009} at room temperature.
Biologically, FMO acts as an energy transmitting wire, delivering an electronic excitation created by photon absorption in the chlorosome antenna to a reaction center where it induces charge separation.
It exists in the form of a trimer with a protein backbone and $3 \times 7$ bacteriochlorophyll-a molecules, each with different transition energies set by the local electrostatic environment.
These pigments and energy transfer pathways are illustrated in Figure \ref{fig:fmo-pathways}.
Several quantitative estimates of the importance of excitonic coherence under particular models suggest that it makes $\sim$10\% contribution to transfer energy transfer efficiency in this system \cite{Rebentrost2008a, Wu2011}.
The section of the energy transfer path from site 1 to 2 is particularly unusual, since it is energetically uphill while these sites also have the strongest electronic coupling of any pair of sites in the complex.
It has been speculated that these factors may indicate a role for quantum coherence in contributing to unidirectional energy flow through this system by avoiding trapping in local minima of the energy landscape \cite{Ishizaki2009a}.
All other steps of the FMO electronic energy transfer pathways in the direction towards the reaction center are energetically downhill, consistent with a pathway that would be optimal for classical energy transport.
A key question that is of special importance for the FMO complex, is thus how the system efficiently directs energy transport away from the chlorosome antenna towards the reaction center. Clearly the overall energy gradient in the system plays a role, but does electronic coherence also facilitate unidirectional energy transfer? More generally, can excitonic coherence assist excitation transfer over the uphill steps found in rough energy landscapes?

A striking feature of all experiments showing electronic quantum beatings in photosynthesis to date is that they have been performed on small sub-units of light harvesting antenna systems, such as the 7 pigments of the FMO complex \cite{Savikhin1997,Engel2007,Panitchayangkoon2010} or 
the 14 pigments in LHC II \cite{Calhoun2009, SchlauCohen:2012wh}. Yet natural light harvesting antennas are typically composed of 
hundreds 
or thousands of pigment molecules organized into many pigment-protein complexes through which energy passes on route to the reaction center \cite{Blankenship2002}.
In addition, natural excitation is by sunlight, not ultrafast laser pulses. 
There is dispute about 
whether or not coherences can arise after excitation by natural light \cite{Cheng2009, Mancal2010, Ishizaki2011, Brumer2011}, but many pigment-protein sub-units are actually more likely to receive excitations indirectly, as a result of weak coupling to another complex in the larger network.
Clearly, understanding the role of excitonic coherence in 
a single protein-pigment complex requires placing the energy transfer 
within and through that system in the context of appropriate initial conditions, as determined by its role in the larger ``supercomplex.''
Accordingly, a second key question for evaluating the relevance of quantum beats to energy transfer on biological scales is whether or not excitonic coherence could be either arise or be maintained in the process of transfer between different sub-units.

In the remainder of this paper we shall address these two open questions with a general theoretical framework employing a novel basis for analyzing the excitonic dynamics of weakly coupled pigment-protein complexes.
First, we develop an analysis of such weakly coupled complexes that suggests a mechanism for how coherence should arise and recur in the process of energy transfer.
We find that coherence can continue to be regenerated during long-range energy transport between weakly linked sub-units of a larger excitonic system.  This is the `propagation' of quantum coherence, whereby a process of continual renewal following incoherent quantum jumps may allow non-zero coherence to last indefinitely, despite rapid decay after each jump.  This has significant implications for long-range energy transfer in light harvesting 
supercomplexes composed of multiple units that individually support coherence, such as photosystems I and II \cite{Blankenship2002}.  Second, we address the question as to whether 
such spatially propagated intra-complex quantum coherence enables unidirectional flow of energy, with a specific example that is inspired by the uphill energetic step in FMO.  We construct an explicitly solvable ratchet model to show that in this situation, the non-equilibrium nature of even limited quantum beating may allow for qualitatively new types of dynamics.  In particular, we show that under biologically plausible conditions, these dynamical features could allow for the operation of a coherently enabled ratchet effect to enhance directed energy flow through light harvesting systems.  These analyses provide new understanding of physical mechanisms reliant on quantum coherence that could be relevant to the function of natural photosynthetic systems.

\section{Spatial propagation of coherence}
\label{sec:prop-coherence}

In photosynthetic energy transfer, excitons typically need to travel through a series of protein-pigment complexes before reaching reaction centers \cite{Blankenship2002}.
To accurately understand the role of coherent dynamics between individual sub-units in such a supercomplex, it is first necessary to understand which particular sub-unit states donate or accept excitations for inter-complex transfer.
As we shall show in this work, the nature of these states informs us whether or not coherence arises under natural conditions in the process of energy transfer.
Moreover, in the context of a light-harvesting complex which is a subcomponent of a larger light-harvesting apparatus, the precise nature of the acceptor states on the complex and the donor states from the complex is paramount to assessing the possible relevance of coherent dynamics in the complex. To draw an analogy to the circuit model of quantum computation \cite{Nielsen2000}, these states serve as effective choices of initial states and measurement basis states, respectively, for dynamics on an individual complex. 
Both of these states need to differ from energy eigenstates in order for strictly unitary dynamics to influence measurement outcomes. The measurement outcomes correspond to observable energy transfer, for which differences in rates or success probabilities could in turn influence biological function.

Since inter-complex couplings are relatively weak, in our analysis we treat them perturbatively, as in multichromophoric generalizations of F\"orster theory used to calculate overall transition rates between donor and acceptor complexes consisting of multiple chromophores \cite{Jang2004, Sumi1999, Scholes2000}.
Our starting point is the equation of motion for the reduced density matrix, which is derived with the following adaptation of the multichromophoric energy transfer rate model \cite{Jang2004}. 
The zeroth order Hamiltonian is $H_0 = H_\D + H_\A$ where $H_{\D} = H^e_{\D} + \sum_{ij} B_{\D_{ij}} \ketbra{\D_i}{\D_j} + H^g_\D$, with $H^e_{\D}$ the electronic Hamiltonian of the donor complex, and corresponding definitions for the acceptor complex $\A$. 
States $\ket{\D_j}$ and $\ket{\A_k}$ for $j=1,\ldots,n$ and $k=1,\ldots,m$ form an arbitrary orthonormal basis for donor and acceptor single-excitation electronic states and
 $B_{\D_{ij}}$ are bath operators that couple the electronic chromophore states to environmental states of the pigment-chromophore system. The ground state donor (acceptor) bath Hamiltonian $H_\D^g$ ($H_\A^g$) can be taken without loss of generality to be a set of independent harmonic oscillators.
We assume that no bath modes are coupled to both the donor and acceptor so that $[H_D,H_A]=0$ \cite{Jang2004}.
The donor and acceptor complexes are coupled by a dipolar interaction
$H_c = \bm{J} + \bm{J}^\dagger$ with $\bm{J}  = \sum_{jk} J_{jk} \ket{\D_j}\bra{\A_k}$.
Calculating the evolution of the reduced system density matrix $\sigma = \tr_{B} \rho$ to second order in $H_c$ 
(\cite{Jang2004} and Appendix \ref{sec:derivation-mc-fret}) yields
\begin{align}
	\frac{\sigma_{k k^\prime}}{dt} &= \sum_{j j^{\prime} k^{\prime\prime} } \frac{J_{j^{\prime} k^{\prime\prime} }}{4 \pi \hbar^2}  \int_{-\infty}^\infty d\omega \Big[ J_{j k} \ E^{j^{\prime} j}_{\D} \! (t,\omega) \ I^{k^{\prime\prime} k^\prime}_{\A} \! (\omega) \notag\\[-6pt]
	&\qquad\qquad\qquad\qquad + J_{j k^\prime} \ E^{j j^{\prime}}_{\D} (t, \omega)  \ I^{k k^{\prime\prime}}_{\A} (\omega) \Big] \label{eq:mcfret-acceptor} \\
	\frac{d\sigma_{j j^\prime}}{dt} &= - \sum_{k k^\prime j^{\prime\prime}} \frac{J_{j^{\prime\prime} k^{\prime}}}{4 \pi \hbar^2} \int_{-\infty}^\infty d\omega \Big[ J_{j k} \ E^{j^{\prime} j^{\prime\prime}}_D \!(t,\omega)\  I^{k k^\prime}_A(\omega) \notag\\[-6pt]
	&\qquad\qquad\qquad\qquad + J_{j^\prime k} \ E^{j^{\prime\prime} j^{\prime}}_D \!(t,\omega)\  I^{k^\prime k}_A(\omega) \Big] \label{eq:mcfret-donor}
\end{align}
for the reduced acceptor and donor density matrix elements, respectively, where $\bm{E}_{\D}(t,\omega)$ and $\bm{I}_{\A}(\omega)$ denote 
matrices of donor and acceptor lineshape functions (see Eqs.~(\ref{eq:acceptorlineshape}--\ref{eq:donorlineshape})).
We emphasize that these results hold for arbitrary system-bath coupling strength, provided that the donor-acceptor coupling is weak: at this point in our analysis we have not yet made any assumption of weak system-bath coupling.

Instead of focusing on the multichromophoric energy transfer rate between complexes that results from summing Eq.~\eqref{eq:mcfret-acceptor} or \eqref{eq:mcfret-donor} over all diagonal terms \cite{Jang2004}, we focus
here on important features relevant to quantum coherence apparent from Eqs.~(\ref{eq:mcfret-acceptor}--\ref{eq:mcfret-donor}) directly.
These equations indicate that acceptor populations $\ketbra{\A_k}{\A_k}$ will  grow and donor populations  $\ketbra{\D_j}{\D_j}$ will decay only
if there is at least one non-zero coupling term $J_{jk} = \bra {\D_j} \bm{J} \ket {\A_k}$ to those states.
Accordingly, we argue that the transfer of electronic states is most sensibly described by the ``inter-complex coupling'' basis in which $\bm{J}$ is diagonal, rather than the site or exciton (energy) basis,
as is assumed in both the original and generalized \cite{Sumi1999, Scholes2000} F\"orster theories.
This inter-complex coupling (ICC) basis is given by the singular value decomposition $\bm{J} = U_\D \tilde{\bm{J}} U_\A^\dagger$, where $\tilde{\bm{J}} $ is a rectangular diagonal matrix and $U_\D$ and $U_\A$ are unitary transformations of donor and acceptor electronic states.
We can thus write the inter-complex coupling as
 $H_c = \sum_l \tilde J_{l} (\ketbra{\tilde D_l}{\tilde A_l} + \ketbra{\tilde A_l}{\tilde D_l})$
in terms of the ICC states $\ket{\tilde D_l} = U_\D \ket{D_l}$ and $\ket{\tilde A_l} = U_\A \ket{A_l}$
for $l \in \{ 1,\ldots, \min(n,m) \}$. 
In the ICC basis, the full electronic Hamiltonian in block-matrix form is
\begin{align}
	\tilde H^e &= \left[\begin{matrix} U_\D^\dagger H_\D^e U_\D & U_\D^\dagger \bm{J} U_\A \\ U_\A^\dagger \bm{J}^\dagger U_\D & U_\A^\dagger H_\A^e U_\A \end{matrix} \right].
\end{align}
Since in general the transformation that diagonalizes $\bm{J}$ will not coincide with the (exciton) eigenbases of $H^e_{D}$ and $H^e_{A}$, population growth of an acceptor ICC state $\ketbra{\tilde A_l}{\tilde A_l}$ thus corresponds to growth of excitonic coherences.

Although in principle Eqs.~(\ref{eq:mcfret-acceptor}--\ref{eq:mcfret-donor}) specify all dynamics relevant to inter-complex transfer, the time-dependent donor lineshape $\bm{E}_{\D}(t,\omega)$ obscures the specific dependence on donor density matrix elements.
Accordingly, we also derive a time-convolutionless quantum master equation (Appendix \ref{sec:weaksys-bath}) under the additional assumption of weak coupling to the bath relative to the donor electronic Hamiltonian $H^e_\D$ \cite{Breuer2002}.
Under this approximation, we see that growth of an acceptor population $\ketbra{\tilde A_{l}}{\tilde A_{l}}$ is proportional to populations only of the coupled donor $\ketbra{\tilde D_{l}}{\tilde D_{l}}$. Likewise, decay of a donor population $\ketbra{\tilde D_{l}}{\tilde D_{l}}$ is proportional to populations only of that donor itself (see Eqs.~(\ref{eq:acceptorchange}--\ref{eq:donorchange}) in Appendix \ref{sec:weaksys-bath}).
Accordingly, inter-complex transfer rates may show oscillations reflecting donor quantum beats, since the ICC states on the donor which transmit excitations do not necessarily correspond to energy eigenstates.
While this part of our argument is only rigorous in the case of weak system-bath coupling, which is not necessarily the case for FMO and other light harvesting systems \cite{Ishizaki2009a}, our simulations find excellent agreement even for moderate strength environmental coupling, as we show below.

To test this analysis of inter-complex energy transfer, we first consider its predictions
for the special case in which there is only one non-zero inter-complex coupling in the ICC basis, $H_c = J_\ast \ketbra{D^\ast}{A^\ast} + h.c$.
If the acceptor is always initialized in the state $\rho_\A^\ast = \ketbra{\A^\ast}{\A^\ast}$, then when back-transfer to the donor is neglected as is valid in the perturbative limit, the acceptor density matrix should be well described by
\begin{align}
	\rho_\A(t) = \int_0^t dt^\prime \frac{d p_A(t^\prime)}{dt^\prime} \mathcal{G}(t-t^\prime) \rho_\A^\ast,
	\label{eq:acceptorpredict}
\end{align}
where $\mathcal{G}(t)$ is the Greens function denoting evolution of the 
acceptor-bath system for time $t$, with the bath initialized at equilibrium.
If the predicted donor state is correct,
then neglecting temporary bath reorganization effects,
the rate of inter-complex transfer should then be proportional to the population of the predicted donor state, for a predicted inter-complex transfer rate
\begin{align}
	\frac{dp_A(t)}{dt} \propto p_{D^\ast}(t), 
	\label{eq:donorpredict}
\end{align}
where $p_{D^\ast}(t)$ denotes the probability of the donor being in the state $\ket{D^\ast}$ and $p_A$ the total probability of the excitation being on the acceptor.

\begin{figure}[] 
	\includegraphics[]{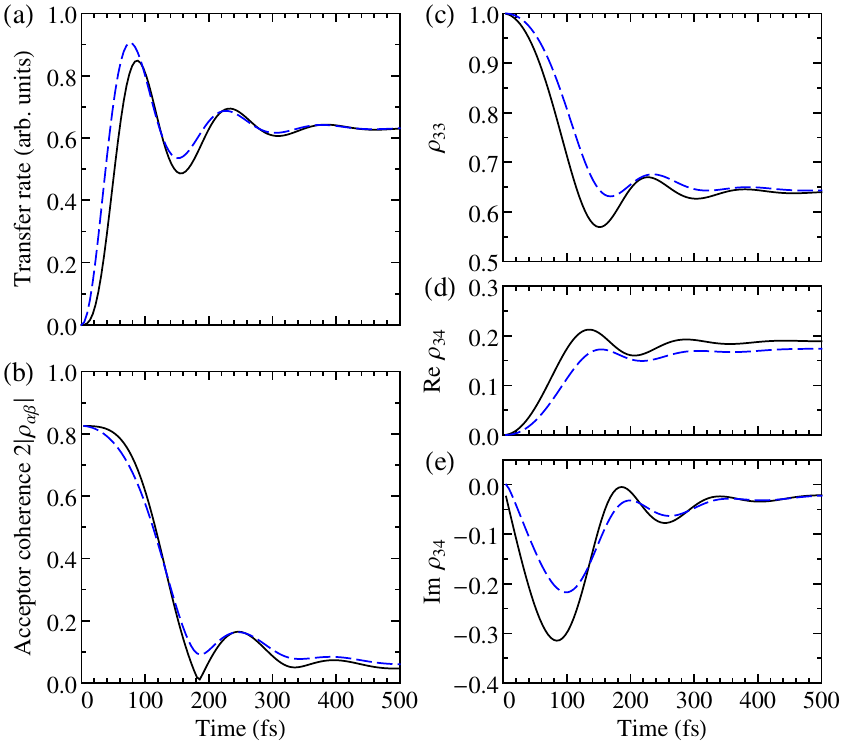}
	\caption{\label{fig:perturbation-error}(Color online) Testing the theory of 
	 propagation of coherence via the inter-complex coupling 
	(ICC) basis. 
	Simulations were made for the coupled dimer model described in the text, with ICC donor and acceptor states $\ket{D^\ast} = \ket 2$, $\ket{A^\ast} = \ket 3$, respectively, and initial condition $\ket{\psi_0} = \ket{1}$.
(a) Simulated (solid black, 2CTNL) and predicted (dashed blue, Eq.~\eqref{eq:donorpredict}) inter-complex transfer rate as a function of time. 
(b) Coherence between the two acceptor excitonic eigenstates $\alpha$ and $\beta$ as a function of time. (c) Population of site 3 in the acceptor, i.e., $\rho_\A^\ast = \ketbra{3}{3}$, as a function of time. (d) Real and (e) imaginary parts of the 3-4 site coherence for the simulated (solid black) and predicted (dashed blue, Eq.~\eqref{eq:acceptorpredict}) acceptor density matrix $\rho_\A$, as a function of time. 
The acceptor density matrix $\rho_\A$ was normalized to unit probability at all times in panels (b-e).
	}
\end{figure}

For a model system, these predictions show remarkable agreement
with results derived from an independent simulation based on 
a 2nd-order cumulant time-nonlocal (2CTNL) quantum master equation
\cite{Ishizaki2009}. 
We consider transfer between two dimer complexes (labelled sites 1, 2 and sites 3, 4), 
with intra-dimer Hamiltonian parameters matching those of the 1-2 dimer of FMO (Appendix \ref{sec:fmo-hamiltonian-svd}, Eq.~\eqref{eq:H_FMO}) and inter-complex coupling $\bm{J} = J_\ast \ketbra{2}{3}$.
We perform calculations in the limit $J_\ast \to 0$ (see Appendix \ref{sec:uni-methods}),
to ensure accuracy of the perturbative description and
eliminate back-transfer effects.  
The 2CTNL calculations are carried out at \SI{300}{K} with a bath modeled by a Debye spectral density with reorganization energy \SI{35}{cm^{-1}} and correlation time $\SI{50}{fs}$, with the initial condition on site 1 \cite{Ishizaki2009a}. 
Figure \ref{fig:perturbation-error} compares simulated 2CTNL results with the predicted time-dependent inter-complex transfer rate, Eq.~\eqref{eq:donorpredict}, and acceptor density matrix, Eq.~\eqref{eq:acceptorpredict} (normalized to unity for greater clarity), calculated from the 2CTNL results.
The results show that estimates based on the dominant elements of the ICC basis provide an accurate representation of both the energy transfer rate (panel a) and acceptor density matrix (panels b-e).
We see that transfer of excitation in the ICC basis from $\ket{D^\ast}$ to $\ket{A^\ast}$ produces a superposition of acceptor eigenstates (of $H_A$) that gives rise to excitonic coherence (panel b) and hence to oscillatory behavior of both the site populations (panel c) and coherences (panels d-e).
Two features are of particular significance, since they show that these ICC-dominated dynamics satisfy the conditions that are necessary for intra-dimer coherence to be relevant to larger scale energy transfer, namely that the dynamics guarantee the preparation and measurement of states which are not energy eigenstates.
The first feature is that the inter-dimer transfer rate clearly tracks coherent oscillations of the donor population $\ketbra{D^\ast}{D^\ast}$ (panel a). The second feature is that the acceptor is initialized in a state with non-zero excitonic coherence (panel b).
Although Fig.~\ref{fig:perturbation-error} shows results for only a single initial condition, additional simulations (not shown) show that these features hold for arbitrary initial conditions of the donor. In particular, excitonic coherence in the acceptor (and thus coherent beating) is triggered even when the initial condition in the donor has no such excitonic coherence.

A simple example of the usefulness of the ICC basis is to determine the initial conditions for electronic excitation transfer through the FMO complex.  The recently discovered 8th chromophore \cite{Ben-Shem2004, Tronrud2009} provides a plausible donor to the remainder of the complex since it sits on the side nearest the chlorosome antenna complex \cite{Busch2011}.
Since structural information concerning the
location of the FMO complex is limited, standard practice to date has been to choose initial and final conditions for simulation of energy transfer in FMO
based on approximate orientation and proximity of chromophores.  The choice of such initial conditions has varied \cite{Adolphs2006, Busch2011, Mohseni2008}, particularly with regards to whether or not the initial quantum states include any excitonic coherence.  Evaluation of the ICC basis between a donor complex consisting solely of site 8 and an acceptor complex consisting of the remainder of the complex (sites 1-7) implies that the acceptor state is mostly localized on site 1 (see Appendix \ref{sec:fmo-hamiltonian-svd}).  We illustrate this in Figure \ref{fig:fmo-pathways}(a). This initial condition is not an energy eigenstate, so the resulting \emph{in vivo} dynamics would necessarily start from a state with coherence in the exciton basis and thus give rise to the quantum beating seen in the laboratory experiments \cite{Engel2007}.

\section{Coherent versus thermal transport in a model dimer with an energy gradient}
\label{sec:modeldimer}

The way in which this coherence regenerating transfer mechanism can yield a non-trivial biological role for coherence  can be already illustrated with a
simple model dimer complex connected to other complexes.
Choosing inter-complex couplings to and from the dimer to be at individual sites as above implies that initialization and transfer in the ICC basis will be at these sites.
Consider preferred `forward' excitation transfer to be that from the donor at site 2 onward to the next complex. Then the asymptotic probability of successful transfer through the complex will be proportional to that population. 
For a dimer, the electronic Hamiltonian is given by
\begin{align*}
	H = 
	 \left[\begin{matrix}\cos \theta&\sin\theta\\-\sin\theta&\cos\theta\end{matrix}\right]
	\left[\begin{matrix}0& 0\\0&\Delta E\end{matrix}\right]
	\left[\begin{matrix}\cos \theta&-\sin\theta\\\sin\theta&\cos\theta\end{matrix}\right],
\end{align*}
where $\theta$ is the mixing angle, which measures the intra-dimer delocalization, and $\Delta E$ the exciton energy difference.  A non-zero mixing angle corresponds to non-zero exciton delocalization, as indicated by the inverse participation ratio $N = 1/(\sin^4\theta + \cos^4\theta)$.

\begin{figure}[] 
	\includegraphics[]{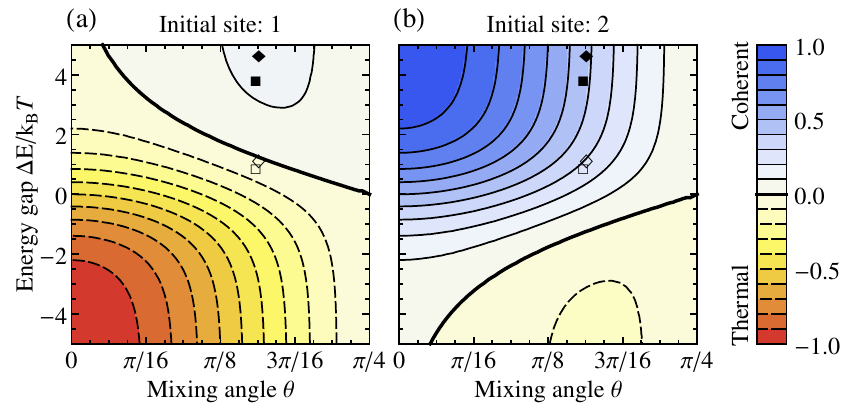}
	\caption{\label{fig:dimer-advantage}(Color online) Difference between coherent and thermal populations on site 2, $p^\text{coh}_{i\to 2} - p^\text{th}_2$, as a function of dimer Hamiltonian parameters for (a) initial site $i=1$ and (b) initial site $i=2$. 
	The empty symbols {\tiny$\square$} and {$\diamond$} indicate location of parameters for the 1-2 dimer in the FMO complex of \emph{C.\ tepidum}
	 at \SI{300}{K} as determined in 
	 Refs.~\citenum{Adolphs2006} and \citenum{Cho2005} respectively. Filled symbols indicate the corresponding parameters at \SI{77}{K}.
	}
\end{figure}

The dimer admits two extreme models of energy transfer: quantum beating due to coherent evolution and instantaneous relaxation to thermal equilibrium between excited electronic states. Instantaneous relaxation provides an upper bound on the speed of excitation transfer governed by a classical master equation, since the dynamics governed by such equations are driven toward thermal equilibrium. This is imposed by the requirement of detailed balance which governs classical dynamics, whether between sites as in F\"orster theory \cite{Laidble1998} or between exciton populations as in variants of Redfield theory \cite{Yang2002}.
For instantaneous relaxation to the thermal distribution, the probability that site 2 is occupied is independent of the initial condition:
\begin{align}
	p^\text{th}_{2} &\propto \bra 2 e^{-\beta H} \ket 2
	= \frac{\cos^2 \theta + e^{\beta \Delta E} \sin^2 \theta}{1 + e^{\beta \Delta E}} \label{eq:p_thermal}.
\end{align}
In contrast, boosts in population due to quantum beating are not restricted by such classical limits \cite{Jang2008, Cheng2007, Ishizaki2009}.
For coherent motion with initialization at site 1, the time-averaged probability of an excitation at site 2 is
\begin{align}
	p^\text{coh}_{1 \to 2} &= \langle \left| \bra 1 e^{-iHt} \ket 2 \right|^2 \rangle_t
	= 2 \cos^2 \theta \sin^2 \theta \label{eq:p_coherent},
\end{align}
while for initialization at site 2, we have $p^\text{coh}_{2 \to 2} = 1 - p^\text{coh}_{1 \to 2}$.
Figure~\ref{fig:dimer-advantage}  plots the difference between coherent and thermal population on site 2, as a function of both the intra-dimer delocalization measure $\theta$ and energy difference $\Delta E$, for both possible initial conditions.  
It is evident that regardless of initial conditions, for a sufficiently uphill energetic arrangement ($\Delta E >0$) intra-dimer quantum beating will be asymptotically more effective than intra-dimer thermalization in enabling transfer onward from the complex via site 2.
The location of the FMO parameters in Fig.~\ref{fig:dimer-advantage} shows that the 1-2 dimer of FMO satisfies such an arrangement at \SI{77}{K} and is on the borderline for strictly enhanced transfer due to coherence at room temperature.

\section{Design for biomimetic %quantum coherent
ratchet}

The asymmetry between incoherent and coherent population transfer seen above for a simple model dimer suggests a design principle that could be exploited for enhanced unidirectional transfer \cite{Ishizaki2009a} and, more generally, a novel type of ratchet based on quantum dynamics \cite{Reimann1997}. Ratchets and Brownian motors \cite{Hanggi2009} utilize a combination of thermal and unbiased non-equilibrium motion to drive directed transport in the presence of broken symmetry.
To take advantage of such a ratchet effect,
strongly linked chromophores with coherent transfer not limited by detailed balance should have an uphill energy step relative to the desired direction of transport, whereas weakly linked chromophores with incoherent transfer steps should be arranged downhill.

\begin{figure}[] 
	\includegraphics[]{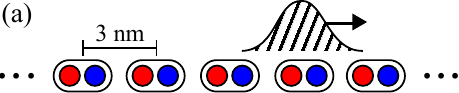}
	\vspace*{0.1in} \\
	\includegraphics[]{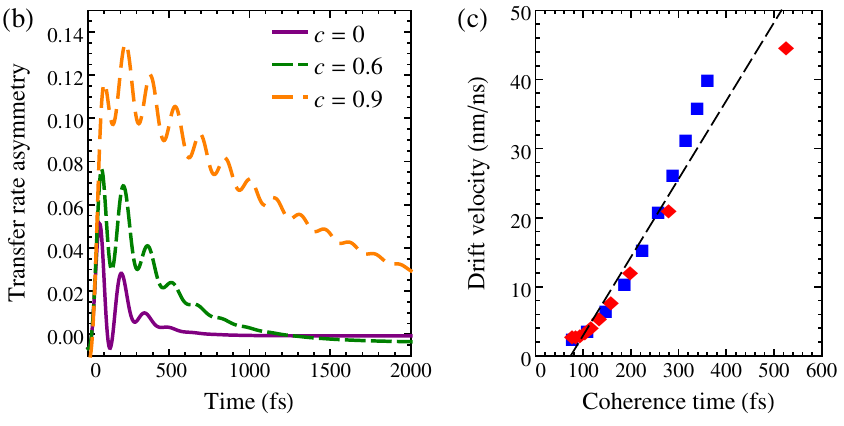}
	\caption{\label{fig:unidirectionality}(Color online) Biased energy transport in an excitonic wire due to spatial propagation of coherence.
	(a) A weakly linked chain of heterodimers is arranged such that the higher energy state is always to the right, with a typical inter-dimer distance of \SI{3}{nm}. The arrow indicates the direction of biased transport.
	(b) Relative asymmetry between left and right inter-dimer transfer rates (Eq.~\eqref{eq:rate-assymetry} of Appendix \ref{sec:uni-methods}) as a function of the time before transferring for a dimer excitation initialized in the asymptotic distribution of site populations.
	(c) Drift velocity vs coherence time as modified by bath correlation time (squares) and cross correlation coefficient between dimer sites (diamonds).
	The dashed line is a linear fit to guide the eye.
	Full details of the simulations in panels b and c are in Appendix \ref{sec:uni-methods}.
	}
\end{figure}

As a proof of principle, we present an example in which this coherent ratcheting effect results in asymptotic spatial bias of transport. Consider a weakly linked chain of heterodimers breaking spatial inversion symmetry, as illustrated in Figure \ref{fig:unidirectionality}(a).
In any classical random walk, transition rates must satisfy detailed balance to assure thermal equilibrium. This guarantees that a classical walk along such a chain is unbiased (Appendix \ref{sec:classicalproof}).
However, we have carried out 2CTNL quantum simulations on small chains of dimers that suggest that including the effects of coherence in each dimer breaks the symmetry of detailed balance to yield a non-zero drift velocity.
Since simulations with the 2CTNL approach would be computationally prohibitive for large numbers of dimers, these simulations were carried out on a chain of three weakly linked dimers with parameters for each dimer matching 
those of the 1-2 dimer of FMO used earlier and an inter-dimer coupling of $\SI{15}{cm^{-1}}$.
We note that this inter-complex coupling strength is well below the
cut-off below which energy transport in light harvesting complexes is
usually described by completely incoherent hopping within
F\"orster theory, though without the possibility of coherence regeneration \cite{Brixner2005, Cho2005}.
The results of these simulations are used to define left and right inter-dimer transition rates for the central dimer. These are then used together with ICC initial conditions from our analysis of the inter-complex coupling as input into a generalized classical random walk describing energy transfer along the chain of dimers. 
The physical model corresponds to the chain shown in panel (a) of Figure \ref{fig:unidirectionality}, with red sites (dimer internal site 1) at energy zero, while  blue sites (dimer internal site 2) are at energy \SI{120}{cm^{-1}}. 
Full details of the construction of transition rates and of the set up and solution of the generalized random walk are described in Appendix \ref{sec:uni-methods}.

Formally, this theoretical approach corresponds to using the microscopic quantum dynamics within and between complexes to define state-specific rates between complexes that generate `quantum state controlled' incoherent energy transfer dynamics over long distances.
A feature of this simulation strategy is that we have \emph{a priori} eliminated the possibility of reaching true thermal equilibrium, since we do not include the long range coherence terms between different dimers. However, it is reasonable to expect the long term influence of such coherences to be negligible, since the inter-dimer couplings are very small. We choose this hybrid approach to the excited state dynamics since we wish to base our simulations on numerically exact calculations, made here with the 2CTNL method that is valid in both limits of strong and weak environmental couplings \cite{Ishizaki2009}.

These quantum state controlled incoherent dynamics can generate a significant bias in the spatial distribution of excitation transfer when the intra-dimer dynamics display long lasting quantum coherence.
We analyzed the random walk with both Monte Carlo simulations on long finite chains and an analytic solution \cite{Fedja2010} of the asymptotic mean and variance of the distribution, as detailed in Appendix \ref{sec:uni-methods}.
Figure \ref{fig:unidirectionality}(b) shows that the underlying asymmetry of transfer rates and violation of detailed balance dynamics is due to the non-equilibrium state of the donor.
The bias is in the forward direction, corresponding to the uphill step within dimers.
Figure \ref{fig:unidirectionality}(c) plots the asymptotic
drift velocity of the random walk against the timescale of excitonic coherence.
We determine this coherence time from a best fit of the timescale of exponential decay of intra-dimer excitonic coherence.
The timescale of coherent oscillations was tuned in two ways, (i) by changing the bath correlation time and (ii) by increasing the spatial correlations between the chromophore-bath couplings \cite{Ishizaki2010a}.
We see a close correlation between the timescale for quantum beating and the magnitude of the bias, regardless of the underlying physical mechanism used to tune the coherence time.
In general, we cannot rule out the possibility that a non-equilibrium/non-Markovian classical model might also yield such biased transport. (We already ruled out such a possibility for a Markovian classical model in Appendix \ref{sec:classicalproof}.)
However, this strong correlation between the duration of quantum beating, independent of its origin, and the asymptotic transport bias supports our interpretation that in this model system the ratchet effect is due to quantum coherent motion. Indeed, the fact that the drift velocity appears to approach a small or zero value as the coherence time goes to zero in Figure \ref{fig:unidirectionality}(c) shows that any contribution deriving from classical non-equilibrium system/non-Markovian bath dynamics here is extremely small relative to that deriving from the quantum coherence maintained in the system degrees of freedom.

Since our results demonstrate a ratchet effect, it is important to consider why this sort of motion is not forbidden by the second law of thermodynamics.
The answer is that the system is never allowed to reach thermal equilibrium along the infinite chain of dimers. This feature is shared by the excitations transferred in natural light harvesting systems, which also do not exist for long enough to reach equilibrium.
The resulting directed motion shows some similarity to the operation of a quantum photocell \cite{Scully2010}, where coherence can (in principle) allow for enhanced conversion of energy by similarly breaking a limit imposed by detailed balance.
In both cases, no additional source of energy is supplied besides that of the non-equilibrium photon which creates the initial excitation. This contrasts with the operation of typical classical and quantum brownian motors \cite{Hanggi2009}, where detailed balance is broken by applying an additional driving force.

Our results for an infinite chain of heterodimers confirm the effectiveness of our ratchet for energy transfer, which we ascribe to the combination of intra-complex excitonic coherence within dimers and an uphill intra-complex energy gradient.
The non-zero drift velocity means that over long distances this ratchet offers a quadratic improvement in transfer times over
any corresponding classical walk, which is unbiased (Appendix \ref{sec:classicalproof}). However, in contrast to the speedup offered by quantum walks \cite{Hoyer2010}, this ratchet requires only short ranged and short lived coherences that will be resilient to the static and dynamic disorder of biological environments.
This spatial bias constitutes a preferential direction for the energy flow across multiple light-harvesting complexes.
It could thus be of direct biological relevance for FMO, which serves as a quantum wire connecting the antenna complex to the reaction center.
We therefore now consider the implications for FMO in more detail.

\section{Role of coherent energy transport in the Fenna-Matthews-Olson complex}

We now specifically consider the role of the coherent dynamics in the uphill step energy of the FMO complex, which corresponds to the 1-2 dimer in the usual notation (see Fig.~\ref{fig:fmo-pathways}).
Since the 1-2 dimer is relatively weakly coupled to the other chromophores in the complex, we may consider transfer to and from this dimer on the basis of our perturbation analysis using ICC states.  
 By performing a singular value decomposition of the appropriate coupling matrices (see Appendix \ref{sec:fmo-hamiltonian-svd}), we find that the
dominant couplings to and from this dimer are
from site 8 to site 1, and from site 2 to site 3.
This suggests the relevance of our dimer model from Section \ref{sec:modeldimer}, where site 8 acts as a donor to the 1-2 dimer, and site 2 in turn acts as a donor to the 3-7 complex. 
\begin{figure}[] 
	\includegraphics[]{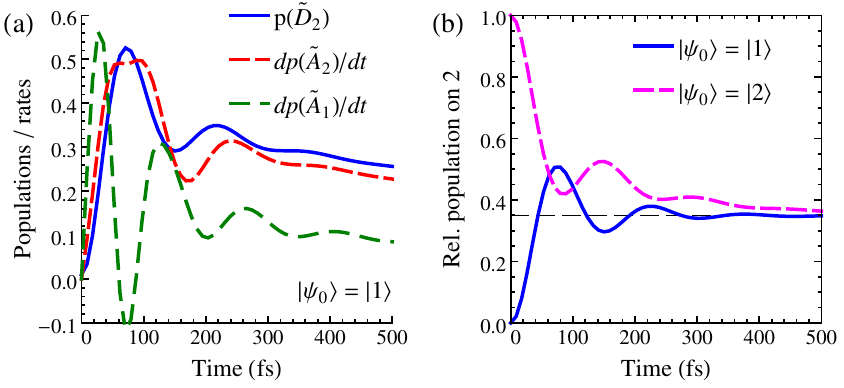}
	\caption{\label{fig:fmo-as-generalized-forster}(Color online) Simulations of FMO dynamics at room temperature.
	(a)
Population of dominant ICC donor state at site 2 in the 1-2 dimer ($\ketbra{\tilde D_2}{\tilde D_2}$, solid line) compared with the rate of population change of its ICC acceptor state in the 3-7 complex ($\ketbra{\tilde A_2}{\tilde A_2}$, dashed line) and the rate of population change of the 
	other ICC acceptor state not coupled to this donor state ($\ketbra{\tilde A_1}{\tilde A_1}$, dash-dotted line), for the initial condition is $\ket{\psi_0} = \ket 1$.. The time derivatives have been scaled to aid comparison of correlations. (b) Population of site 2 relative to the total 1-2 dimer population, $p_2/(p_1+p_2)$, for both choices of initial conditions.}
\end{figure}
Figure \ref{fig:fmo-as-generalized-forster}(a) presents results of a 2CTNL simulation on sites 1-7 of FMO, partitioning FMO between donor state on the 1-2 dimer and acceptor states on the remaining sites 3-7 (i.e., neglecting the prior donation from the 8th site to the 1-2 dimer).
The corresponding ICC donor/acceptor states are given in Appendix \ref{sec:fmo-hamiltonian-svd} 
(Eqs.~(\ref{eq:fmo-ICC-1}--\ref{eq:fmo-ICC-2})).
We see that the ICC donor population $\ketbra{\tilde D_2}{\tilde D_2} \approx \ketbra{2}{2}$ is positively correlated with the rate of growth of its coupled ICC acceptor $\ketbra{\tilde A_2}{\tilde A_2}$, but negatively correlated with the growth of the other ICC acceptor state $\ketbra{\tilde A_1}{\tilde A_1}$, to which it is not coupled.  This is in agreement with the predictions of our theory from Section \ref{sec:prop-coherence}.
(The small deviations arise because FMO is not quite in the regime of validity for the perturbation theory and because Eq.~\eqref{eq:donorpredict} is not strictly valid for the situation with two acceptor states $\ket{\tilde A}$.) 
The simulation is performed at \SI{300}{K} for a bath correlation time of \SI{50}{fs} as described previously \cite{Ishizaki2009a}.

As indicated by the location of the FMO 1-2 dimer Hamiltonian parameters in Figure \ref{fig:dimer-advantage}, this particular chromophore dimer appears be optimized to have an uphill energy gradient just large enough so that
excitonic coherence enhances transfer if initialized at site 2 (panel a) without also suppressing transfer initialized at site 1 (panel b). In Figure \ref{fig:fmo-as-generalized-forster}(b) we compare the portion of dimer population on site 2 from 2CTNL calculations with the classical upper bound of the thermal average, for initial conditions in both of the ICC states $\ket 1$ and $\ket 2$.
The populations show quantum beating deriving from partially coherent motion.
The populations also agree with the predications of our simple dimer model (Section \ref{sec:modeldimer}): the population at site 2 averaged over quantum beats (Eq.~\eqref{eq:p_coherent}) is nearly equal to the thermal average (Eq.~\eqref{eq:p_thermal}) for initialization at site 1 and greater than the thermal average for initialization at site 2.

As evident from Fig.~\ref{fig:fmo-as-generalized-forster}, these temporary boosts in population at site 2 due to coherence should correspond to enhanced biological function, since they drive excitations preferentially toward site 3 instead of backwards toward the antenna complex. 
The 1-2 dimer in the FMO complex thus appears to act as one link of our proposed ratchet for energy transport, thereby enhancing unidirectional energy flow through this pathway of the FMO complex.
Consistent with previous numerical estimates of the contribution of coherent energy transfer to transfer efficiency in photosynthetic systems \cite{Rebentrost2008a, Wu2011}, we expect that any quantitative enhancement to the speed of energy transfer through FMO due to such a limited ratchet effect will be relatively small compared to the contribution of incoherent energy relaxation.
Refining such estimates is not the purpose of this work.
Rather, our new dynamical model of transport between ICC states allows us to propose specific physical advantages that the electronic coherence provides for general light harvesting systems.
In particular, we established the ability to propagate excitonic coherence between weakly coupled sub-units and to use the ratchet effect to enhance unidirectional transport.
It is also conceivable that the cumulative contribution of many such small contributions from propagating coherence through the entire photosynthetic apparatus of green sulphur bacteria (of which FMO is only a small part) could indeed make a major contribution to the speed of energy transfer, as in the full ratchet example.

\section{Conclusions}

We have proposed a microscopic mechanism for the propagation of excitonic coherence in energy transfer between photosynthetic complexes. 
The mechanism allows coherence to be propagated between sub-units of a large light harvesting ``supercomplex'' that is composed of multiple complexes that individually support coherence. Our analysis shows that the key role in the inter-complex transfer is played by the inter-complex coupling (ICC) basis, rather than  energy or site bases employed by prior analyses.
By utilizing ICC donor and acceptor states, we showed that coherence can enable biased energy flow through a ratchet mechanism.  We provided evidence that this same principle acts to ensure unidirectional energy flow in the FMO complex.
Since one-way transmission of electronic energy from the antenna complex to the reaction center constitutes the main function of the FMO protein in the light harvesting apparatus of green sulfur bacteria, this supports a biological role for the electronic quantum coherence in this particular light harvesting system.

Propagating coherence provides both a mechanism by which coherent motion can influence transfer rates and photosynthetic efficiency in light harvesting systems of arbitrary size (a possible quantum advantage), and a scalable method for multi-scale modeling of such excitonic systems without neglecting the contributions of coherence (practical simulations).
Our example and analysis of a coherently enabled ratchet effect along a chain of heterodimers demonstrates both of these features.
This proof of principle model shows that even short-lived excitonic coherence can, since it propagates spatially, lead to large scale dynamics that are incompatible with any completely classical description. 
We also demonstrated how fully quantum models need only to be used for tightly coupled sub-complexes, while transfer between sub-complexes may take the form of classical hops with connections between states of different sub-complexes constrained by the ICC basis.

Similar techniques should allow us to assess the role of coherence in natural photosynthetic super-complexes with hundreds of chlorophyll molecules, such as arrays of LH1 and LH2 rings in purple bacteria, and the photosystem I and II super-complexes of higher plants \cite{Blankenship2002}.
For example, direct calculation of ICC states should help us evaluate the significance of long lasting coherences in bacterial reaction centers \cite{Lee2007, Lee2009}, since these systems are also usually unlikely to absorb light directly \cite{Blankenship2002} and thus might be benefitting from recurrence of coherence propagated from light harvesting complexes.
Some bacterial reaction centers also feature an uphill step opposing the direction of desired energy flow \cite{Timpmann1995}, resembling the energetic arrangement in the uphill step of the FMO complex.

Finally, the dynamics in our chain of heterodimers model constitute a novel type of ratchet that utilizes spatial propagation of quantum coherence in place of a driving force and as such are also of more general interest. Thus, in addition to excitonic systems, we expect that a similar ratchet effect could be demonstrated in other experimental systems described by spin-boson Hamiltonians, such as cold atoms in optical lattices \cite{Sebby-Strabley2006,Recati2006}.

%TC:break Acknowledgments
\begin{acknowledgments}
We thank MathOverflow user ``fedja'' for providing the analytical analysis of our random walk \cite{Fedja2010} presented in Appendix \ref{sec:uni-methods} and Yuan-Chung Cheng for discussions.
This work was supported in part by DARPA under award N66001-09-1-2026 and by the U.S.\ Department of Energy under contracts DE-AC02-05CH11231 and DE-AC03-76SF000098.
S.H.\ is a D.O.E.\ Office of Science Graduate Fellow. 
\end{acknowledgments}

\appendix

\section{Extending multichromophoric F\"orster theory}
\label{sec:derivation-mc-fret}

In this Appendix, we provide additional technical details of the derivation of Eqs.~(\ref{eq:mcfret-acceptor}--\ref{eq:mcfret-donor}).
Consider a system under zeroth order Hamiltonian $H_0$ with perturbation Hamiltonian $V$.
In the interaction picture $\rho_\text{I}(t) = e^{i H_0 t/\hbar} \rho(t) e^{-i H_0 t/\hbar}$ the von Neumann equation is
\begin{align}
	\frac{d\rho_\text{I}}{dt} = -\frac{i}{\hbar}[V_\text{I}(t), \rho_\text{I}(t)],
	\label{eq:von-neumann}
\end{align}
which can be formally integrated to yield,
\begin{align}
	\rho_\text{I}(t) = \rho_\text{I}(0) - \frac{i}{\hbar} \int_0^t dt^\prime [V_\text{I}(t^\prime), \rho_\text{I}(t^\prime)].
	\label{eq:von-neumann-integrated}
\end{align}
Inserting Eq.~\eqref{eq:von-neumann-integrated} into Eq.~\eqref{eq:von-neumann}, keeping all terms second order in $V$
and transforming back to the Schr\"oedinger picture
yields the second order contribution to the time-convolutionless equation of motion
\begin{align}
	\frac{d\rho}{dt} &= -\frac{1}{\hbar^2} \int_0^t d\tau\ [V, [e^{-i H_0 \tau/\hbar} V e^{i H_0 \tau/\hbar}, \rho(t)]],
	\label{eq:perturbgeneral}
\end{align}
where we approximated $e^{-i H_0 \tau/\hbar} \rho(t-\tau) e^{i H_0 \tau/\hbar} \approx \rho(t)$ (valid to this order in $V$).

We are interested in the lowest order contribution to the donor and acceptor electronic states from a perturbative treatment of the inter-complex coupling terms with the zeroth order Hamiltonian given by that of the otherwise independent donor and acceptor complexes. Since a first order treatment of $H_c$ gives coherences between the donor and acceptor but no population transfer, we thus consider the second order contribution.
Note that our perturbation parameter is this inter-complex coupling $H_c$ rather than the usual coupling to the bath, so our results will apply to any bath coupling strength.
Substituting our perturbation $V = H_c$ into Eq.~\eqref{eq:perturbgeneral} and tracing over the bath yields the equation of motion for acceptor states,
\begin{align}
	&\frac{d\sigma_{k k^\prime}}{dt} = \sum_{j} \sum_{j^{\prime\prime} k^{\prime\prime}}  \frac{J_{j^{\prime\prime} k^{\prime\prime}}}{\hbar^2} \int_0^t d\tau \notag \\[-3pt]
	& \times \tr_\text{B}\! \Big[  J_{j k} \bra{D_j} \rho_D^e \rho_A^g e^{-i H_0 \tau/\hbar} \ket{D_{j^{\prime\prime}}} \bra{A_{k^{\prime\prime}}} e^{i H_0 \tau/\hbar} \ket{A_{k^\prime}} \notag \\[0pt]
	& + \bra{A_k} e^{-i H_0 \tau/\hbar} \ket{A_{k^{\prime\prime}}}
	\bra{D_{j^{\prime\prime}}} e^{i H_0 \tau/\hbar} \rho_D^e \rho_A^g \ket{D_{j}} J_{j k^\prime} \Big],
	\label{eq:acceptor-explicit}
\end{align}
where we used the initial condition $\rho = \rho_D^e \rho^g_A$ for a general excited donor state with the acceptor in the ground state at thermal equilibrium \cite{Jang2004}.
The donor equation is similar and thus omitted for conciseness.
To simplify these equations, we use the following identity, which follows from the substitution $H_0 = H_A + H_D$, by employing the cyclic properties of the trace as well as the assumptions that the donor and acceptor baths are independent and that all strictly donor and strictly acceptor terms commute,
\begin{align}
	&\tr_\text{B}\! \Big[ \bra{D_j} \rho_D^e \rho_A^g e^{-i H_0 \tau/\hbar} \ket{D_{j^{\prime\prime}}} \bra{A_{k^{\prime\prime}}} e^{i H_0 \tau/\hbar} \ket{A_{k^\prime}} \Big] \notag\\
	&\qquad= \tr_\text{B}\! \Big[ e^{i H_D^g \tau/\hbar} \bra{D_j} \rho_D^e e^{-i H_D \tau/\hbar} \ket{D_{j^{\prime\prime}}} \Big] \notag \\[-3pt]
	&\qquad\qquad \times \tr_\text{B}\! \Big[ e^{-i H^g_A \tau/\hbar} \bra{A_{k^{\prime\prime}}} e^{i H_A \tau/\hbar} \ket{A_{k^\prime}} \rho_A^g \Big].
\end{align}
Substitution of this identity and its Hermitian conjugate into Eq.~\eqref{eq:acceptor-explicit} gives a form amenable to substitution by products of acceptor and donor lineshape functions \cite{Jang2004}, given by
\begin{align}
	I^{k^\prime k}_A (\omega) =& \int_{-\infty}^{\infty} dt \, e^{i \omega t} \notag \\
	&\times \tr_\B  \! \left\{ e^{i H^g_A t/\hbar} \bra{A_{k^\prime}} e^{-i H_A t/\hbar} \ket{A_k} \rho^g_A \right\} \label{eq:acceptorlineshape} \\
	E^{j^\prime j}_D (t, \omega) =& 2 \int_{0}^t dt^\prime e^{-i \omega t^\prime} \notag \\
	\times& \tr_\B \! \left\{ e^{-i H^g_D t/\hbar} \bra{D_{j^\prime}} e^{i H_D t/\hbar} \rho^e_D \ket{D_j} \right\}.
	\label{eq:donorlineshape}
\end{align}
Inserting these lineshapes yields Eqs.~(\ref{eq:mcfret-acceptor}--\ref{eq:mcfret-donor}).
Explicit dependence upon $t$ in the donor lineshape can be removed by applying the Markov approximation, that is, allowing the upper limit of this integral to be extended to infinity and assuming that the donor $\rho_D^e$ is stationary. 
This would give rate expressions corresponding to those of equilibrium multichromophoric F\"orster theory \cite{Jang2004}.

We note that the result in Eqs.~(\ref{eq:mcfret-acceptor}--\ref{eq:mcfret-donor}) shows that these equations do not necessarily conserve positivity, a feature hidden by the sum over states to determine an overall transfer rate \cite{Jang2004}. This is 
an intrinsic limitation of the perturbative approach
to inter-complex transfer. In particular, these expressions may predict the creation of non-physical acceptor coherences of the form $\ketbra{A_k}{A_{k^\prime}}$ even without necessarily increasing both of the corresponding population terms $\ketbra{A_k}{A_k}$ and $\ketbra{A_{k^\prime}}{A_{k^\prime}}$. For this reason, in determining ICC states, we explicitly only consider those states which will experience population growth or decay.

\section{Weak system-bath coupling}
\label{sec:weaksys-bath}

Under an approximation of weak system-bath coupling relative to the electronic Hamiltonian of the isolated donor $H_D^e$, the full density matrix can be factorized in the form $\rho(t) = \rho^\B_\text{eq} \sigma(t)$ between the equilibrium state of the bath $\rho^\B_\text{eq} = \rho^g_D \rho^g_A$ and the electronic state of the system $\sigma(t)$. We do not need to assume weak system-bath coupling for the acceptor, since it is already in factorized form in the electronic ground state. Likewise, we do not need to assume weak system-bath coupling relative to the inter-complex coupling $H_c$, since $H_c$ does not enter into lineshape expressions for either the donor or the acceptor.
Accordingly, Eq.~\eqref{eq:perturbgeneral} becomes,
\begin{align}
	\frac{d\sigma}{dt} &= -\frac{1}{\hbar^2} [V, [K(t), \sigma]], \label{eq:fullmaster}\\
	K(t) &= \int_0^t d\tau \tr_\B [e^{-i H_0 \tau/\hbar} V e^{i H_0 \tau/\hbar} \rho^\B_\text{eq}],
\end{align}
where the explicit perturbation $V$ is still the inter-complex coupling $H_c$.
The Markov approximation is given by taking $t \to \infty$, in which case we write $K = \lim_{t\to\infty} K(t)$. Since $V$ and $K$ are not in general equal, the Markovian expression is not in Lindblad form and thus does not necessarily conserve positivity, a standard limitation of perturbative derivations of quantum master equations  \cite{Breuer2002}.

For convenience, from now on we apply the Markov approximation. Similar conclusions hold in the more general case.
We then can write Eq.~\eqref{eq:fullmaster} in terms of the evolution of each density matrix element as
\begin{align}
	\frac{d\sigma_{ab}}{dt}	&= \frac{1}{\hbar^2} \sum_{cd} R_{abcd} \sigma_{cd}
	\label{eq:redfield-sum}
\end{align}
by defining Redfield-like tensor elements
\begin{align}
	R_{abcd} &= -\sum_{e} \left[ \delta_{db} V_{ae} K_{ec} + \delta_{ac} K_{de} V_{eb} \right] \notag \\[-6pt]
	&\qquad\qquad\qquad\qquad + K_{ac} V_{db} + V_{ac} K_{db}. \label{eq:redfield-like}
\end{align}
To evaluate our particular model of inter-complex transfer, it is useful to define an acceptor lineshape that only depends upon the bath state in the same form as the donor lineshape (Eq.~\eqref{eq:donorlineshape}),
\begin{align}
	\mathcal{E}^{j^\prime j}_D (\omega) &= \int_{-\infty}^\infty dt\ e^{i \omega t} \notag \\
	&\ \times \tr_\B \left\{ e^{i H^g_D t/\hbar} \bra{D_{j^\prime}} e^{-i H_D t/\hbar} \ket{D_j} \rho^{g}_D \right\}. \label{eq:donorlineshape2}
\end{align}
For weak system-bath coupling and in this Markov limit, we can write the general donor lineshape $\bm{E}_\D (t,\omega)$ (Eq.~\eqref{eq:donorlineshape}) in terms of this modified lineshape,
\begin{align}
	E_\D^{j^\prime j}(\infty, \omega) &= \sum_{j^{\prime\prime}} \mathcal{E}^{j^\prime j}_\D (\omega) \sigma_{j j^{\prime\prime}}.
	\label{eq:modified-donor-lineshape-relation}
\end{align}
Since $K$ is Hermitian, to evaluate the model of Section~\ref{sec:prop-coherence} it suffices to calculate $K_{jk} = \bra{D_j} K \ket{A_k}$. Using the cyclic property of the trace and inserting the donor and acceptor lineshapes, we find
\begin{align}
	K_{jk} &= \frac{1}{4 \pi} \sum_{j^\prime k^\prime} J_{j^\prime k^\prime} \int_{-\infty}^\infty d\omega \ \mathcal{E}^{j^\prime j}_D (\omega)  I^{k^\prime k}_A (\omega).
	\label{eq:K-matrix-elements}
\end{align}

We can now evaluate the tensor elements in Eq.~\eqref{eq:redfield-like} that specify the influence of donor density matrix elements on inter-complex transfer, either by using the matrix elements for $K$ given in Eq.~\eqref{eq:K-matrix-elements} or by using the equivalence in Eq.~\eqref{eq:modified-donor-lineshape-relation} to insert the modified donor lineshape into Eqs.~(\ref{eq:mcfret-acceptor}--\ref{eq:mcfret-donor}).
The relevant tensor elements for the change of the acceptor elements due to the donor are given by,
\begin{align}
	R_{k k^\prime j_0 j_0^\prime} &= \sum_{k^{\prime\prime} j} \frac{J_{j k^{\prime\prime}}}{4 \pi}  \int_{-\infty}^\infty d\omega \Big[ J_{j_0^\prime k^\prime} \ \mathcal{E}^{j_0 j}_D \! (\omega)  \ I^{k k^{\prime\prime}}_A \!(\omega) \notag \\[-6pt]
	&\quad\qquad\qquad + J_{j_0 k} \ \mathcal{E}^{j j_0^\prime }_D \! (\omega) \ I^{k^{\prime\prime} k^\prime }_A \! (\omega) \Big]
	\label{eq:acceptorchange}
\end{align}
and for the change of the donor itself due to donating an excitation,
\begin{align}
	R_{j j^\prime j_0 j^{\prime}_0} 
	&= - \sum_{k k^\prime j^{\prime\prime}} \frac{J_{j^{\prime\prime} k^\prime}}{4\pi} \int_{-\infty}^\infty d\omega \,  \Big[\delta_{j^\prime j^{\prime}_0} J_{jk} \ \mathcal{E}^{j_0 j^{\prime\prime}}_D (\omega)  I^{k k^\prime}_A (\omega) \notag \\
	&\quad\qquad\qquad + \delta_{j j_0} J_{j^\prime k} \ \mathcal{E}^{j^{\prime\prime} j_0^\prime}_D (\omega)  I^{k^\prime k}_A (\omega)   \Big].
	\label{eq:donorchange}
\end{align}
Since these tensor elements are given in terms of an arbitrary basis for the donor and acceptor electronic states, we may write them in terms of the ICC states for which $\bm{J}$ is diagonal. Considering the elements that affect populations ($k = k^\prime$ for acceptor, $j = j^\prime$ for donor), it is then evident that in the ICC representation, the factors $J_{j_0 k}$ restrict nonzero contributions from donor states $j$ to only those deriving only from the coupled donor ICC state $\ketbra{D_{j_0}}{D_{j_0}}$.
Each of these terms also includes a sum over other inter-complex coupling matrix elements and lineshapes, but these only affect the magnitude of the allowed transitions.
In the case where there is only a single nonzero ICC coupling, Eq.~\eqref{eq:redfield-sum} thus reduces to Eq.~\eqref{eq:donorpredict} of the main text.

Note that since our donor and acceptor lineshapes (Eq.~\eqref{eq:donorlineshape2} and Eq.~\eqref{eq:acceptorlineshape}) take identical forms under weak environmental coupling, resulting forward and backward transfer rates will be equal and thus may not necessarily respect detailed balance. Therefore we do not calculate actual rates using Eq.~\eqref{eq:donorlineshape2}.

\section{Proof that classical Markovian transport is unbiased}
\label{sec:classicalproof}

Consider a classical Markov process that models transport along a chain of 
dimers like the chain we used for the quantum coherent ratchet model (Figure \ref{fig:unidirectionality}). We impose only one requirement on the transition rates in this model: thermal equilibrium must be a steady state. For a Markov process, this implies that the transition rates satisfy detailed balance. On each dimer, we consider two states in the single excitation subspace, which could equally well be sites or excitons.
For simplicity, consider excitation transfer only between nearest-neighbor states (similar symmetry constraints guarantee unbiased transport even in the general case).
Then an excitation initially at the lower energy state of each dimer has two possible moves: with probability $p$ to the higher energy state of the same dimer, or with probability $1-p$ to the higher energy site of the neighboring dimer to the left (the non-zero coupling guarantees that eventually the excitation will move). For an excitation at the higher energy state of a dimer, detailed balance requires that the rate of transitions to each neighboring state (at the lower energy) is the rate from those states scaled by the Boltzmann factor $e^{\beta \Delta E}$.
The relative intra- vs inter-dimer transition rates are still the same, so the probability of an intra-dimer jump is still $p$, and $1-p$ for the inter-dimer jump, now to the right. Since every jump alternates between high and low energy states, and these intra- and inter-dimer transitions are alternatively to the left and to the right, on average the random walk must be stationary.

\section{Ratchet methods}
\label{sec:uni-methods}

\paragraph{Propagation of coherence}

With weak coupling between different dimers, inter-dimer transfer should follow the principles of our theory of propagated coherence described in Section \ref{sec:prop-coherence}.
Here we restrict the inter-dimer coupling to be between nearest neighbors, to simplify the singular value decomposition. This is a reasonable approximation for realistic dipole-dipole couplings in light harvesting arrays because the $1/r^3$ scaling ensures a rapid fall-off with inter-chromophore distance.
Accordingly, an ICC analysis tells us that after an inter-dimer transfer the dynamics will be reset with the initial condition on the site in the dimer nearest the side from which the excitation was received. Thus if an excitation is received from the dimer to the left (right), we restart dynamics the initial condition is on the left (right) site of the new dimer.
For the complex consisting of the $n$th dimer,
this corresponds to the explicit inter-complex coupling matrix (in the ICC basis) 
\begin{align}
	\bm{J} = \sum_{\epsilon\in\{-1,+1\}} J \ketbra{n,\epsilon}{n-\epsilon,-\epsilon},
	\label{eq:ICC-chain}
\end{align}
where $J$ is an arbitrary inter-dimer coupling strength and $\ket{n, \epsilon}$ is the state corresponding to occupation of the right ($\epsilon = -1$) or left ($\epsilon = +1$) site of dimer $n$.
We also assume that upon excitation at a site in a new dimer, the baths of the donating chromophore will instantaneously revert to thermal equilibrium.
Accordingly, since the chain of dimers is periodic, we can build overall dynamics in this manner from full quantum calculation of just four time-dependent transfer rates corresponding to left or right transfer to a neighboring dimer from each of the two initial conditions on sites.

\paragraph{Scaled 2CTNL calculations}
For computational feasibility, we based our calculations of transfer rates on scaling the results of 2CTNL simulations on a three dimer (six site) system.
We may denote the left, central and right dimers as $-1, 0, +1$, respectively.
We use the 2CTNL method because it accurately models dynamics under both strong and weak environmental coupling \cite{Ishizaki2009}.
Since we need to calculate rates neglecting back-transfer while these simulation methods describe full system dynamics, we calculate the dynamics here with the inter-dimer coupling $J_0$ set to be very small so that back-transfer was negligible.
To begin, we need cumulative transition probabilities $F_{\epsilon\delta}^0(t)$ for the transition from initial conditions $\epsilon \in \{+1,-1\}$ to neighboring dimer $\delta \in \{+1,-1\}$ at time $t$.
From our simulations on the three dimer chain, the quantity $F_{\epsilon\delta}^0(t)$ is simply the total population at time $t$ on dimer $\delta$ obtained by starting with initial condition on site $\epsilon$ of 
the central dimer.  The probability density as a function of $t$, which is the transition rate, is then given by $f_{\epsilon\delta}^0 (t) = \frac{\partial}{\partial t} F_{\epsilon\delta}^0(t)$ and evaluated numerically.
We then scale the transfer rate to obtain the rescaled rate $f_{\epsilon\delta} (t) = (J/J_0)^2 f_{\epsilon\delta}^0 (t)$ corresponding to the coupling $J$. The rescaled cumulative transition probability is obtained by numerical integration, $F_{\epsilon\delta} (t) = \int_0^t d\tau f_{\epsilon\delta} (\tau)$.
This scaling of the transfer rate assumes that to lowest order in $J$ the transfer rate is proportional to $J^2$, as given by Eq.~\eqref{eq:mcfret-acceptor}.
For our parameters, we found that the scaled transfer rate $f_{\epsilon\delta}(t)$ does indeed converge as $J_0 \to 0$ and that using a value $J_0 = \SI{1}{cm^{-1}}$ was sufficiently small to make any error negligible.
This method neglects the effects of excitation loss on the dynamics of the donating dimer, which is reasonable to lowest order in $J$.
We simulated the three dimer chain using two such 2CTNL calculations, one for each initial condition on a specific site of the central dimer. 
Calculations including spatially correlated baths on each dimer were performed as described previously \cite{Ishizaki2010a}.
In principle, one could perform calculations taking into account static disorder, but we do not expect static disorder would influence our qualitative findings since the primary effect of disorder is to limit delocalization and our scaling procedure already constrains exciton delocalization to individual dimers.

\begin{figure}[]
	\includegraphics[]{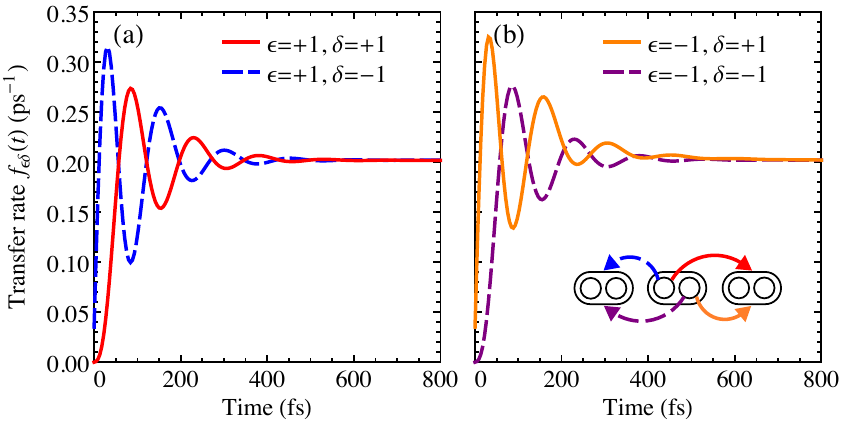}
	\caption{\label{fig:uni-transfer-rates}(Color online) Inter-complex transfer rates $f_{\epsilon\delta}(t)$ with initial conditions (a) $\epsilon=+1$ and (b) $\epsilon = -1$ for steps $\delta = +1$ (solid lines) and $\delta = -1$ (dashed lines). These rates are derived from 2CTNL simulations for a chain of three dimers with correlation time \SI{50}{fs} and no spatial correlations, as described in the text.
	The transfer rates oscillate, corresponding to quantum beatings in the donor, but eventually converge to the same equilibrium rate, as required to satisfy detailed balance. However, at early times, the left ($\delta = +1$) and right ($\delta = -1$) transfer rates 
	are not equal, oscillating out of phase.  When averaged over the limiting distribution $\pi_\epsilon$ of the initial condition this gives rise to the marked short time asymmetry in the left and right inter-dimer transfer rates that is shown in Figure \ref{fig:unidirectionality}(b) of the main text. This asymmetry, although small in absolute terms, is amplified by being repeated over many hops between dimers and is responsible for the asymptotic bias of the random walk.
	}
\end{figure}

Figure \ref{fig:uni-transfer-rates} gives an example of these transfer rates $f_{\epsilon\delta}(t)$. The transfer rates oscillate, corresponding to quantum beatings in the donor, but eventually converge to the same equilibrium rate, as required to satisfy detailed balance. However, at early times, the left ($\delta = +1$) and right ($\delta = -1$) transfer rates 
are not equal, oscillating out of phase.  When averaged over the limiting distribution $\pi_\epsilon$ of the initial condition this gives rise to the marked short time asymmetry in the left and right inter-dimer transfer rates that is shown in Figure \ref{fig:unidirectionality}(b) of the main text. This asymmetry, although small in absolute terms, is amplified by being repeated over many hops between dimers and is responsible for the asymptotic bias of the random walk.

\paragraph{Generalized random walk}
With each transition to a neighboring dimer only depending on the initial conditions at each dimer, the dynamics constitute a type of Markov chain controlled random walk known as a semi-Markov process \cite{Schuss2010}.
In each step of the random walk, we start with a state of the form $\ket{n, \epsilon}$ denoting occupation of the right ($\epsilon = -1$) or left ($\epsilon = +1$) site of the $n$th dimer.
The cumulative transition probabilities $F_{\delta\epsilon}(t)$ for transitioning from $\ket{n, \epsilon}$ to 
the left or right dimer ICC acceptor state
$\ket{n+\delta, \delta}$ for $\delta = \pm 1$
(see Eq.~\eqref{eq:ICC-chain})
are determined by the rescaled 2CTNL calculations as described above.
The update $\epsilon^\prime = \delta$ is the initial condition for the excitation on the new dimer following from our ICC analysis.
The walk is memory-less in terms of a two-dimensional clock variable $(n,t)$ denoting  ``space-time'' position, but the initial condition at each dimer nevertheless functions as an additional ``coin'' degree of freedom $\epsilon$ that controls the likelihood of jumps to a new clock state $(n^\prime, t^\prime)$.

\paragraph{Monte-Carlo algorithm}
Given the transition probabilities $F_{\epsilon\delta}(t)$ for this random walk, we used two techniques to calculate the 
long time behavior of the overall random walk. First, we performed Monte-Carlo simulations of the evolution for a total time $T$ by averaging over trajectories of many jumps.
We start by setting the clock to the state $(n,t) = (0,0)$ and the coin 
to $\epsilon = +1$.
We sample from the distribution of possible space-time shifts $\xi_{\epsilon\delta} = (\delta, t_{\epsilon\delta})$ by choosing a pair $(u_1, u_2)$  of independent uniformly distributed random numbers between 0 and 1. For convenience, we define the final transition probability $p_{\epsilon\delta} \equiv \lim_{t \to\infty} F_{\epsilon\delta}(t)$. If $u_1 \leq p_{\epsilon, +1}$, we choose $\delta=+1$ for this jump; otherwise, $\delta=-1$. The time $t_{\epsilon\delta}$ it takes for this jump is determined by numerically solving the equation $u_2 = F_{\epsilon\delta}(t_{\epsilon\delta})/p_{\epsilon\delta}$ for $t$. We then update the clock $(n^\prime, t^\prime) = (n + \delta, t + t_{\epsilon\delta})$ and the coin $\epsilon^\prime = \delta$. This process is repeated until time $t + t_{\epsilon\delta} > T$, at which point we 
record the location of the previous dimer $n$ as the final state of that trajectory. The probability density of the final distribution over dimers is 
derived by binning over many such trajectories ($\sim$5000).
Empirically, our Monte-Carlo simulations suggest that the distribution converges to a normal distribution characterized by its mean and variance, as expected from a central limit theorem for weakly dependent variables \cite{Durrett1996}.

\paragraph{Analytic model}
Second, we calculated the mean and variance of final distribution analytically in the asymptotic limit of the total walk time $T\to\infty$, using the method suggested in Ref.~\citenum{Fedja2010}. Since the results of these calculations agreed with the Monte-Carlo simulations but were much faster, we use this second method for the plots in Figure \ref{fig:unidirectionality}.
To begin, we calculate the moments of the transition time $t_{\epsilon\delta}$ for each jump $\epsilon \to \delta$,
\begin{align}
	\E(t_{\epsilon\delta}) &= \frac{1}{p_{\epsilon\delta}} \int_0^\infty t f_{\epsilon\delta}(t) dt \\
	\E(t_{\epsilon\delta}^2) &= \frac{1}{p_{\epsilon\delta}} \int_0^\infty t^2 f_{\epsilon\delta}(t) dt
\end{align}
by numerical integration.
Now, note that transitions between coin states can be described as a Markov chain with transition matrix $P$ with entries given by the final transition probabilities $p_{\epsilon\delta} = \lim_{t\to\infty} F_{\epsilon\delta}(t)$ as defined above, i.e.,
\begin{align}
	P = \left( \begin{matrix} p_{+1,+1} & p_{+1,-1} \\
	p_{-1,+1} & p_{-1,-1} \end{matrix} \right).
	\label{eq:transitionmatrix}
\end{align}
Accordingly, the limiting distribution $\pi_\epsilon$ over the coin space is given by the left eigenvector of $P$ with eigenvalue 1, that is, the solution $\pi$ of the equation $\pi_{\epsilon^\prime} = \sum_{\epsilon} \pi_\epsilon p_{\epsilon\epsilon^\prime}$.
The quantity $\pi_\epsilon p_{\epsilon\delta}$ gives the probability of the step $\epsilon \to \delta$ in the limiting distribution.
Recalling the definition of the space-time shift $\xi_{\epsilon\delta} = (\delta, t_{\epsilon\delta})$ associated with the step $\epsilon\to\delta$, 
we obtain an average space-time shift $\bar\xi_{\epsilon\delta} = (\delta, \operatorname E(t_{\epsilon\delta}))$ for this step.
Since the coin will converge to the limiting distribution $\pi_\epsilon$, we now obtain the average space-time shift over all steps as
\begin{align}
	\bar\xi = \E(\xi) = \sum_{\epsilon \delta} \pi_\epsilon p_{\epsilon\delta} \bar\xi_{\epsilon\delta} \equiv (\bar n, \bar t).
\end{align}
Now let $n_T$ denote the spatial position of the random walk after a total time $T$.
This random walk is the sum of $T/\bar t$ independent steps on average, each of which has an average spatial shift $\bar n$. Since the expectation adds linearly, we then obtain the average position of the overall walk as
\begin{align}
	\E(n_T) = \bar n T/ \bar t.
	\label{eq:uni-mean}
\end{align}
Figure \ref{fig:unidirectionality} plots the corresponding drift velocity $v = \E(n_T)/T$.

Given that our random walk appears to converge to a normal distribution, we can fully characterize the distribution with its mean, 
calculated above, and its variance. As a practical matter, the variance indicates the width of the distribution and 
thus determines whether or not a non-zero drift velocity would be observable experimentally.
To calculate the variance, we consider two sources of space-time deviations,
\begin{align}
	\eta_{\epsilon\delta} &= \xi_{\epsilon\delta} - \bar \xi_{\epsilon\delta}  \\
	\mu_{\epsilon\delta} &= \bar\xi_{\epsilon\delta} - \bar\xi,	\label{eq:mu_stepaverage}
\end{align} 
corresponding to deviations $\eta_{\epsilon\delta}$ of the space-time shift of a particular transition from its average 
value, and deviations $\mu_{\epsilon\delta}$ of the average space-time shift for a particular transition from the average space-time shift over all transitions.
Since successive steps are weakly correlated by the ICC conditions,
the latter quantity must be averaged over all possible steps in all possible trajectories.  We therefore define $\bar\mu$ as the single step average obtained by summing $\mu_{\epsilon\delta}$ over all possible sequential steps:
\begin{align}
	\bar\mu = \lim_{n\to\infty} \frac{1}{n} \sum_{i=1}^n \mu_{\epsilon_i \delta_i}.
\end{align}
Note that for a standard Markov chain with no correlation between steps, this single step average reduces to $\mu$. 
Assuming the covariation between $\eta_{\epsilon\delta}$ and $\bar\mu$ is small, we can combine them to calculate the overall space-time covariance matrix,
\begin{align}
	\Var(\xi) &= \Var(\xi - \bar \xi) \\
	&= \Var(\eta + \bar\mu) \\
	&= \Var(\eta) + \Var(\bar\mu). \label{eq:var-xi}
\end{align}
Averaging over the limiting distribution $\pi_\epsilon$, we find for the first contribution to the variance,
\begin{align}
	\Var(\eta) &= \sum_{\epsilon\delta} \pi_\epsilon p_{\epsilon\delta} \Var(\eta_{\epsilon\delta})
	\label{eq:var-eta}
\end{align}
where
\begin{align}
	\Var(\eta_{\epsilon\delta}) = \left( \begin{matrix} 0 & 0 \\ 0 & \Var(t_{\epsilon\delta}) \end{matrix} \right),
\end{align}
with $\Var(t_{\epsilon\delta}) = \E(t_{\epsilon\delta}^2) - \E(t_{\epsilon\delta})^2$.
Now consider 
the second contribution to the variance, $\Var(\bar \mu)$.
The variance of the single step average $\bar\mu$, Eq.~\eqref{eq:mu_stepaverage}, introduces a double sum over products of deviations $\mu_{\epsilon \delta}$.  With a correlated random walk, the products of deviations at different space-time values are also correlated and hence evaluation of these requires enumeration of all possible jumps connecting them, 
where these are determined by the transition probability matrix $P$, Eq.~\eqref{eq:transitionmatrix}. This enumeration,
which constitutes a multi-state generalization of the variance for 
weakly dependent 
processes \cite{Durrett1996}, is given explicitly by
\begin{align}
	\Var(\bar\mu) &= \sum_{\epsilon\delta} \pi_\epsilon p_{\epsilon\delta} \mu^\mathrm{T}_{\epsilon\delta} \mu_{\epsilon\delta} \notag \\
	+& \sum_{\epsilon\delta\rho\sigma} \sum_{m\geq 0} \pi_\epsilon p_{\epsilon\delta} p^{(m)}_{\delta\rho} p_{\rho\sigma} [\mu_{\epsilon\delta}^\mathrm{T} \mu_{\rho\sigma} + \mu_{\rho\sigma}^\mathrm{T} \mu_{\epsilon\delta} ],
	\label{eq:var-mu}
\end{align}
where $p_{\delta\rho}^{(m)} = (P^m)_{\delta\rho}$
and we sum each variable $\epsilon, \delta, \rho, \sigma$ over $\pm 1$. The second term in Eq.~\ref{eq:var-mu} sums up all contributions that $m$-steps apart, where these are specified by the Chapman-Kolmogorov equation \cite{Breuer2002}.
We note that for convenience, instead of explicitly performing the sum over $m$, one can equivalently replace the term $\sum_m p^{(m)}_{\delta\rho}$ in the equation above with $Q_{\delta\rho}$  \cite{Fedja2010}, where
$Q = (1 - P^\ast)^{-1}$ and $P^\ast$ is the non-equilibrium portion of $P$, that is, with entries $p^\ast_{\epsilon\delta} = p_{\epsilon\delta} - \pi_\epsilon$.
Combining Eqs.~\eqref{eq:var-eta} and \eqref{eq:var-mu} into \eqref{eq:var-eta} yields a space-time covariance matrix $\Var(\xi)$ for the 
two dimensional shift variable $\xi$. 
This covariance matrix has explicit entries,
\begin{align}
	\Var(\xi) &= \begin{pmatrix} \Var(n) & \Cov(n,t) \\ \Cov(n, t) & \Var(t) \end{pmatrix}.
\end{align}
To calculate the final spatial variance $\Var(n_T)$, we must take into account the uncertainty associated with the number $m$ of discrete hops that happened in time $T$, in addition to the uncertainty over $n$. To correctly incorporate both contributions, we calculate the variance of the spatial displacement $n$ over a single coin shift over the full two-dimensional coin space,
\begin{align}
	\Var_{(n,t)}(n) &= \Var_{(n,t)}(n - \bar n t /\bar t) \notag \\
	&= \Var(n) - 2 (\bar n / \bar t) \Cov(n, t) + (\bar n / \bar t)^2 \Var(t) \notag \\
	&= (1, -\bar n/\bar t) \Var(\xi) (1, -\bar n/\bar t)^\mathrm{T},
\end{align}
where in the first step we subtracted the average value of $n$ over the coin space.
We write the variances over the full coin space $(n,t)$ to emphasize that they are distinct from terms like $\Var(n)$, which is only over the spatial degree of freedom $n$.
Since the variance adds linearly over $m \approx T/\bar t$ independent steps, we obtain the variance of the distribution after time $T$ as
\begin{align}
	\Var(n_T) &= (1, -\bar n/\bar t) \Var(\xi) (1, -\bar n/\bar t)^\mathrm{T} \frac{T}{\bar t}.
	\label{eq:uni-variance}
\end{align}
The diffusion coefficient for the walk is then given as $D = \Var(n_T)/2T$.

\begin{figure}[t] 
	\includegraphics[]{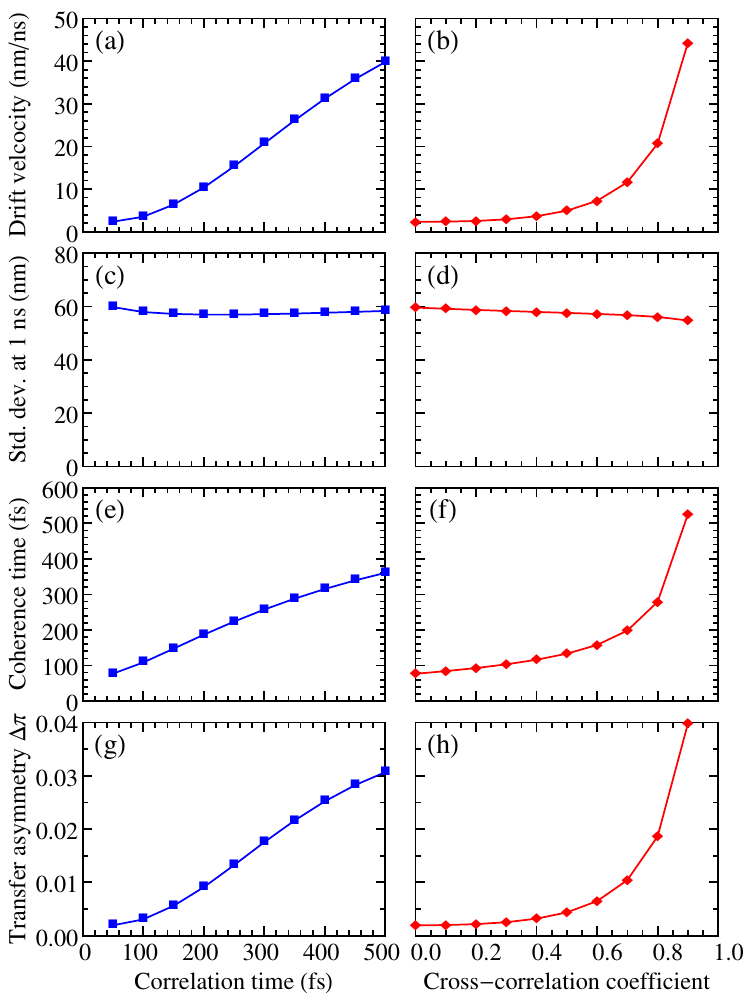}
	\caption{\label{fig:uni-full}(Color online) Full results of simulations used for analysis of the unidirectional random walk, as used to create Figure \ref{fig:unidirectionality} of the main text. The parameters of each dimer match that of sites 1-2 in FMO, as described in the main text. 
	We used a Debye spectral density with a reorganization energy of \SI{35}{cm^{-1}} and variable bath and spatial correlations, as indicated on the figure.
	For the left panels we have variable time correlation and no spatial correlations. For the right panels we vary the spatial correlation 
	(see Ref.~\cite{Ishizaki2010a}) and fix the correlation time at \SI{50}{fs}.
	(a,b) Drift velocity, from Eq.~\eqref{eq:uni-mean}. (c,d) Standard deviation $\sigma = \sqrt{\Var(n_T)}$ of the walk at \SI{1}{ns}, from Eq.~\eqref{eq:uni-variance}. (e,f) Coherence time $\tau$, from a least squares fit of the exponential decay of excitonic coherence, $|\rho_{e_1 e_2}| \sim A e^{-t/\tau} + B$: 
	we evaluate this here and in Figure \ref{fig:unidirectionality} for $t > \SI{100}{fs}$ to exclude non-exponential decay. (g,h) Asymptotic transfer asymmetry $\Delta\pi = \pi_{+1} - \pi_{-1}$ indicating the overall preference for right over left transfer.
	}
\end{figure}

Figure \ref{fig:uni-full} plots the full results of scans over correlation time and cross-correlation coefficients used to create Figure \ref{fig:unidirectionality} of the main text.
We see that the width of the excitation transfer distribution is approximately constant over all parameter choices at about \SI{60}{nm} after \SI{1}{ns}, and that the asymmetry between initial conditions $\Delta \pi = \pi_{+1} - \pi_{-1}$ accounts for most of the variation in drift velocity.
Figure \ref{fig:unidirectionality}(b) of the main text is a plot of the relative transfer rate asymmetry for the limiting distribution of the initial condition $\pi_\epsilon$,
\begin{align}
	A(t) = \frac{\sum_{\epsilon\delta} \pi_\epsilon \delta f_{\epsilon\delta}(t)}{ \sum_{\epsilon\delta} \pi_\epsilon f_{\epsilon\delta}(t)},
	\label{eq:rate-assymetry}
\end{align}
where the sums are over $\epsilon,\delta \in \{-1, +1\}$ as usual.

\section{FMO Hamiltonian and singular value decompositions}
\label{sec:fmo-hamiltonian-svd}

The FMO complex exists in a trimer arrangement, where each monomer contains 7 bacteriochlorophyll molecules,
and three additional BChl molecules (termed the 8th Bchl
for each of the three monomers) are each located between a distinct pair of monomers \cite{Tronrud2009}.
In this paper, we use a Hamiltonian for a monomer of the FMO complex of \textit{C.\ tepidum} calculated by Adolphs and Renger \cite{Adolphs2006}, augmented by dipole-dipole couplings to the 8th BChl site calculated using their same methodology with the crystal structure of Tronrud \emph{et al.}\ \cite{Tronrud2009}.  
We assign each of the three BChl 8 pigments to the monomers with which they have the strongest dipole-dipole coupling.
The Hamiltonian matrix is given below in units of cm$^{-1}$ above \SI{12210}{cm^{-1}},
where elements of the matrix are indexed according to site from 1 to 8:
\begin{align}
	{\small
	\begin{bmatrix}
	 200 & -87.7 & 5.5 & -5.9 & 6.7 & -13.7 & -9.9 & 37.5 \\
	 -87.7 & 320 & 30.8 & 8.2 & 0.7 & 11.8 & 4.3 & 6.5 \\
	 5.5 & 30.8 & 0 & -53.5 & -2.2 & -9.6 & 6. & 1.3 \\
	 -5.9 & 8.2 & -53.5 & 110 & -70.7 & -17. & -63.3 & -1.8 \\
	 6.7 & 0.7 & -2.2 & -70.7 & 270 & 81.1 & -1.3 & 4.3 \\
	 -13.7 & 11.8 & -9.6 & -17. & 81.1 & 420 & 39.7 & -9.5 \\
	 -9.9 & 4.3 & 6. & -63.3 & -1.3 & 39.7 & 230 & -11.3 \\
	 37.5 & 6.5 & 1.3 & -1.8 & 4.3 & -9.5 & -11.3 & ?
	\end{bmatrix}
	}.
	\label{eq:H_FMO}
\end{align}
The energy of site 8 is marked with a question mark to indicate that it is unknown, since it has not been calculated. Accordingly, our simulations of the full FMO complex use only the portion of this Hamiltonian for sites 1-7, as in previous studies \cite{Ishizaki2009a}.

To determine donor and acceptor ICC states for a given coupling matrix $\bm{J}$, we perform the singular value decomposition $\bm{J} = U_\D \tilde{\bm{J}} U_\A^\dagger = \sum_l \tilde{J}_l \ketbra{\tilde D_l}{\tilde A_l}$ as described in Section~\ref{sec:prop-coherence}. Here are the results of two examples we use with our FMO Hamiltonian.
Let the notation $\bm{J}_D^A$ denote the coupling matrix from the donor (D) rows and the acceptor (A) columns of Eq.~\eqref{eq:H_FMO}.
As plotted in Figure \ref{fig:fmo-pathways}, for the coupling from site 8 to sites 1-7, we have $\bm{J}_{8}^{1\text{-}7} = J_\ast \ketbra{D^\ast}{A^\ast}$ with
\begin{align}
	J_\ast = 41.9 \quad
	\ket{D^\ast} = \ket{8} \quad
	\ket{A^\ast} = \begin{bmatrix} -0.912 \\ -0.158 \\ -0.031 \\ 0.043 \\ -0.105 \\ 0.229 \\ 0.275 \end{bmatrix},
\end{align}
with entries $\braket{i}{A^\ast}$ for states $\ket i = \ket 1, \ldots, \ket 7$.
With only a single donor site, the ICC acceptor (donor) state from the singular value decomposition is as simple as the normalized vector corresponding to the dipole-dipole matrix.
Since occupation probabilities correspond to these amplitudes squared, the acceptor among sites 1-7 is mostly (83\%) on site 1, as shown in Fig.~\ref{fig:fmo-pathways}(a).

A less trivial example is given by considering the 1-2 dimer as a donor to and acceptor from the remainder of the complex 3-8. In this case, we have $\bm{J}_{1\text{-}2}^{3\text{-}8} = \sum_{i=1,2} \tilde J_i \ketbra{\tilde D_i}{\tilde A_i}$ with
\begin{align}
	\tilde J_1 = 43.6 \quad
	\ket{\tilde D_1} = \begin{bmatrix} -0.969 \\ -0.247 \end{bmatrix} \quad
	\ket{\tilde A_1} = \begin{bmatrix} -0.297 \\ 0.085 \\ -0.153 \\ 0.238 \\ 0.196 \\ -0.887 \end{bmatrix}, \\
	\tilde J_2 = 34.3 \quad
	\ket{\tilde D_2} = \begin{bmatrix} 0.247 \\ -0.969 \end{bmatrix} \quad
	\ket{\tilde A_2} = \begin{bmatrix} -0.832 \\ -0.274 \\ 0.028 \\ -0.432 \\ -0.193 \\ 0.089 \end{bmatrix},
\end{align}
with entries corresponding to states $\ket i$ in ascending order.
Thus the coupling in the ICC basis is mostly from site 8 (78\%) to site 1 (94\%), and from site 2 (94\%) to site 3 (69\%).
If we omit site 8 from the acceptor, these acceptor and donor states are modified as follows:
\begin{align}
	\tilde J_1 = 19.7 \quad
	\ket{\tilde D_1} = \begin{bmatrix} 0.995 \\ 0.099 \end{bmatrix} \quad
	\ket{\tilde A_1} = \begin{bmatrix} 0.433 \\ -0.257 \\ 0.342 \\ -0.633 \\ -0.479 \end{bmatrix}, \label{eq:fmo-ICC-1} \\
	\tilde J_2 = 34.4 \quad
	\ket{\tilde D_2} = \begin{bmatrix} 0.099 \\ -0.995 \end{bmatrix} \quad
	\ket{\tilde A_2} = \begin{bmatrix} -0.876 \\ -0.254 \\ -0.001 \\ -0.381 \\ -0.153 \end{bmatrix}. \label{eq:fmo-ICC-2}
\end{align}
We use these ICC states in Fig.~\ref{fig:fmo-as-generalized-forster} since only sites 1-7 are included in the 2CTNL simulation, as the site energy of the 8th BChl is unknown (see Eq.~\eqref{eq:H_FMO}), and as it is furthermore unclear whether this 8th BChl is present in all cases in the natural system \cite{Tronrud2009}.

% \bibliography{library}

\end{document}